\documentclass[conference]{IEEEtran}
\IEEEoverridecommandlockouts
\usepackage{cite}
\usepackage{amsmath,amssymb,amsfonts}
\usepackage[ruled,norelsize,vlined,linesnumbered]{algorithm2e}
\usepackage{multirow}
\usepackage{graphicx}
\usepackage{textcomp}
\usepackage{xcolor}

\usepackage{url}
\usepackage{cite}
\usepackage{hyperref}
\usepackage{array}
\usepackage{textcomp}
\usepackage{stfloats}
\usepackage{verbatim}
\usepackage{subfigure}

\def\BibTeX{{\rm B\kern-.05em{\sc i\kern-.025em b}\kern-.08em
    T\kern-.1667em\lower.7ex\hbox{E}\kern-.125emX}}
\begin{document}
	
\title{DeepBurning-MixQ: An Open Source Mixed-Precision Neural Network Accelerator Design Framework for FPGAs\\
\vspace{-0.7em}
}


\makeatletter
\newcommand{\linebreakand}{%
\end{@IEEEauthorhalign}
\hfill\mbox{}\par
\mbox{}\hfill\begin{@IEEEauthorhalign}
}
\makeatother

\author{
	\IEEEauthorblockN{Erjing Luo$^{1,2}$\IEEEauthorrefmark{1}\thanks{\IEEEauthorrefmark{1} These authors contributed equally to this work.}, Haitong Huang$^{1,3}$\IEEEauthorrefmark{1}, Cheng Liu$^{1}$\IEEEauthorrefmark{2}\thanks{\IEEEauthorrefmark{2} Corresponding authors.}, Guoyu Li$^{1,3}$, Bing Yang$^{4}$, Ying Wang$^{1}$, Huawei Li$^{1}$, Xiaowei Li$^{1}$}
	\IEEEauthorblockA{
		$^{1}$SKLP, Institute of Computing Technology, Chinese Academy of Sciences, Beijing, China
	}
	\IEEEauthorblockA{
		$^{2}$School of Information and Electronics, Beijing Institute of Technology, Beijing, China
	}
	\IEEEauthorblockA{
		$^{3}$Dept. of Computer Science, University of Chinese Academy of Sciences, Beijing, China
	}
	\IEEEauthorblockA{
		$^{4}$Dept. of Computer Science and Technology, Harbin University of Science of Technology, Harbin, China
	}
	\IEEEauthorblockA{\{huanghaitong21s, liucheng, liguoyu21s\}@ict.ac.cn, 1120192664@bit.edu.cn}
	\vspace{-2.8em}

 \thanks{This work is supported by the National Key R\&D Program of China under Grant (2022YFB4500405), and the National Natural
Science Foundation of China under Grant 62174162.}
}

\maketitle

\begin{abstract}
	Mixed-precision neural networks (MPNNs) that enable the use of just enough data width for a deep learning task promise significant advantages of both inference accuracy and computing overhead. FPGAs with fine-grained reconfiguration capability can adapt the processing with distinct data width and models, and hence, can theoretically unleash the potential of MPNNs. Nevertheless, commodity DPUs on FPGAs mostly emphasize generality and have limited support for MPNNs especially the ones with lower data width. In addition, primitive DSPs in FPGAs usually have much larger data width than that is required by MPNNs and haven’t been sufficiently co-explored with MPNNs yet. To this end, we propose an open source MPNN accelerator design framework specifically tailored for FPGAs. In this framework, we have a systematic DSP-packing algorithm to pack multiple lower data width MACs in a single primitive DSP and enable efficient implementation of MPNNs. Meanwhile, we take DSP packing efficiency into consideration with MPNN quantization within a unified neural network architecture search (NAS) framework such that it can be aware of the DSP overhead during quantization and optimize the MPNN performance and accuracy concurrently. Finally, we have the optimized MPNN fine-tuned to a fully pipelined neural network accelerator template based on HLS and make best use of available resources for higher performance. Our experiments reveal the resulting accelerators produced by the proposed framework can achieve overwhelming advantages in terms of performance, resource utilization, and inference accuracy for MPNNs when compared with both handcrafted counterparts and prior hardware-aware neural network accelerators on FPGAs. 
\end{abstract}


\IEEEpeerreviewmaketitle

\section{Introduction}

Quantization is a straightforward yet effective approach \cite{dorefanet} \cite{GoogleQAT} to compress neural network models that are usually both computing- and memory-intensive. Since neural network's sensitivity to quantization varies across the layers, mixed-precision quantization that allows more fine-grained quantization achieves significant advantages in both computing efficiency and memory access efficiency compared to classical uniform quantization \cite{HAO}\cite{EnergyEfficientNAS}, which contributes to the neural network processing throughput and energy efficiency eventually. Nevertheless, most of the commodity neural network computing engines including CPU, GPU, and NPU usually have limited support for arbitrary mixed-precision neural network (MPNN) processing especially low data width processing due to the lack of native mixed precision computing elements. In contrast, FPGAs with fine-grained reconfiguration capability can provide model specific implementation \cite{DNNBuilder} \cite{DNNExplorer} and suit various computing requirements of MPNNs with native hardware, and hence, can unleash the potential of MPNNs.

Despite the fine-grained reconfiguration capability, FPGAs mainly rely on digital signal processing unit (DSP) cores with limited reconfiguration for efficient arithmetic implementations. For example, Xilinx integrates DSP48E2 which can support a 27 $\times$ 18 two's complement multiplication on its UltraScale FPGAs\cite{XilinxDSP48E2}, while Intel Arria 10 devices have DSP cores that can be configured to two 18 × 19 multipliers or one 27 × 27 multiplier \cite{IntelDSP}. Although Lookup-Tables (LUTs) can also be utilized to implement arithmetic operations with arbitrary data width, DSPs generally outperform the LUT-based implementations in terms of both latency and energy efficiency \cite{DSP-Packing} especially for higher data width operations. While MPNNs typically involve many low data width operations that are much smaller than data width of primitive DSPs, straightforward implementation of MPNNs can result in considerable waste of the DSP resources. To address the problem, a variety of works have attempted to pack multiple low data width operations into a single DSP block \cite{HiKonv} \cite{XilinxINT4} \cite{XilinxINT8} and make full use of the computing capability of the DSPs. For instance, Xilinx's INT8\cite{XilinxINT8} and INT4\cite{XilinxINT4} demonstrate that two 8-bit multiplications and four 4-bit multiplications can be fit into a single DSP48E2. The authors in \cite{HiKonv} also explored the use of high data width processing engines for low data width processing. 

While MPNN quantization affects not only the accuracy but also the DSP packing efficiency and performance eventually, performing the quantization and DSP packing separately will lead to sub optimal results. In fact, there have been intensive efforts devoted to optimize the neural network model accuracy and performance at the same time. For instance, \cite{HAQ} employs a RL agent with direct hardware metrics feedback to co-optimize accuracy, latency, and energy consumption. \cite{EDD} incorporates quantization and other neural architecture parameters into a design space, and has gradient-decent method for optimizing both algorithms and hardware implementation.
However, there is still a lack of co-optimization between DSP packing and MPNN quantization. 

In this work, we propose a systematic DSP packing algorithm that can squeeze multiple low data width arithmetic operations into a single primitive DSP of FPGAs. While the DSP packing efficiency varies substantially across the convolution kernels with different parameters, the overall performance of a MPNN network can be inconsistent with the data width of the model. Then, we leverage a differentiable NAS to take both the DSP packing and MPNN quantization into consideration and co-optimize the model accuracy and performance at the same time. Finally, with the optimized DSP packing and quantization determined by the NAS, we have a MPNN accelerator generated automatically based on a fully pipelined neural network accelerator template. 


The major contributions of this work can be summarized as follows:

\begin{itemize}
	\item We propose a mixed DSP packing algorithm that takes advantage of both \emph{Kernel Packing} strategy and \emph{Filter Packing} strategy for arbitrary low data width convolution on FPGAs. In addition, we also enhance the DSP packing with \emph{Overpacking} technique and \emph{Operation Separation} technique. The resulting DSP packing algorithm outperforms all the existing strategies significantly.
 
	\item On top of the proposed DSP packing algorithm, we propose a MPNN accelerator design framework that leverages a differentiable NAS to take both the DSP packing and quantization into a consideration and optimizes the model accuracy and performance at the same time. With the optimized DSP packing strategy and quantization setups, this framework can further generate high-performance MPNN accelerator based on a fully pipelined HLS template. The framework is open sourced on Github\footnote{\url{https://github.com/fffasttime/AnyPackingNet/}}.
 
	\item According to our experiments on a set of different MPNNs, the MPNN accelerators generated with the proposed framework outperform state-of-the-art counterparts in terms of performance, resource utilization, and prediction accuracy significantly.
\end{itemize}




\section{Related Work}
Mixed-precision neural networks (MPNNs) that enable just enough data width for each different neural network layer can greatly reduce the requirements of computing, memory bandwidth, and storage. Therefore, MPNNs promise great computing efficiency when compared to neural network models with unified data width. Since the number of low bit-width operations is inconsistent with the realistic computing efficiency due to the lack of primitive MPNN implementation on existing computing engines including CPUs, GPUs, and NPUs. Many prior work seek to co-optimize the quantization and computing efficiency \cite{FracBits} \cite{Once-for-All} to ensure efficient MPNN implementation on specific computing engines. Specifically, \cite{HAQ} \cite{BatchQuant} \cite{allyouneed} \cite{Rethinking} \cite{GPUNAS} \cite{chen2020you} leverage network architecture search (NAS) technique that automates the neural network design for the co-optimization. For instance, \cite{GPUNAS} has a hardware performance model added to differentiable NAS such that performance of neural network candidates on GPUs can be evaluated with the model accuracy at the same time. Similar approaches have also been successfully applied to lightweight neural network design for mobile phones\cite{FBNet}. However, the above hardware aware neural network design and optimization approaches essentially adapt the neural network models to the target computing engines that have fixed computing architectures and have little native low data width operation support, and hence, fail to fully unleash the potential of MPNNs. 

In contrast, FPGAs with fine-grained reconfiguration capability are suitable for model specific customization and can be a good fit for MPNNs. To explore the reconfiguration capability of FPGAs for efficient neural network specific implementation, the authors in \cite{HAO} \cite{HAQ} \cite{EDD} proposed different co-optimization approaches that take both neural network accuracy and hardware implementation efficiency into consideration in a unified framework. \cite{HAQ} applies time-consuming reinforcement learning NAS. Although such NAS approach could have the accelerator design parameters included in the same NAS search space along with the network architecture and have different design metrics considered at the same time during NAS evaluation stage, it enlarges the search space dramatically and usually induce many time-consuming evaluation of metrics such as accuracy, hardware overhead, and implementation quality, which makes the entire optimization prohibitively expensive and difficult to converge. \cite{EDD} formulates the co-optimization as a differentiable NAS problem in terms of both accuracy and implementation efficiency. It avoids the conventional iterative NAS search procedures and reduces the optimization to a standard training procedure. Essentially, it makes the hardware-aware optimization much easier to converge. However, it is difficult to explicitly represent and model all hardware design parameters within a differentiable NAS framework, as many hardware parameters are discrete, making it difficult to construct differentiable proxy loss and obtain global optimal solutions. Different from these above NAS approaches, \cite{HAO} takes the co-optimization as a \textit{integer programming} problem by introducing an accuracy predictor and a performance predictor, which do not rely on any complex black box simulation or evaluation procedures. Particularly, it demonstrates the great potential of using a simplified co-optimization framework for neural network acceleration on FPGAs, but the accuracy prediction can be relatively limited to some specific scenarios. 

In addition, the fine-grained reconfiguration capability of FPGAs has not been explored sufficiently. FPGAs mainly rely on primitive DSP blocks with fixed data width for high performance arithmetic operations while straightforward implementation of low data width such as 2-bit and 3-bit operations on these primitive DSP blocks can lead to considerable waste because the data width of these DSP blocks is much larger. LUTs in FPGAs can also be utilized to construct operations with arbitrary data width, but the performance is usually much lower especially for operations with larger data width due to the more complex routing. To fully explore the fine-grained reconfiguration capability of FPGAs, researchers from both academia and industry have proposed a variety of approaches to pack multiple low-precision operations into a single DSP (especially multiplier) concurrently to fully utilize the primitive DSPs \cite{DSP-Packing} \cite{HiKonv} \cite{XilinxINT4} \cite{XilinxINT8} \cite{WSQ-AdderNet} and then extract the outputs of the low-precision operations from the disjoint bit segments of the DSP blocks concurrently. For instance, Xilinx INT4 optimization \cite{XilinxINT4} leverages the $27\times 18$ two's complement multiplier to simultaneously calculate four 4-bit products. \cite{Interconnect_Aware} \cite{FTDL} proposes to construct efficient low data width matrix-matrix multiplication overlay based on primitive DSP blocks. Particularly, it takes the interconnection between primitives into consideration for the sake of higher operation frequency. Then, it has neural network models implemented based on the overlay to ensure high-performance neural network processing on FPGAs. Essentially, it optimizes the hardware implementation first without being aware of the models and adapts the model to the FPGA implementation afterwards. HiKonv \cite{HiKonv} specifically explores the DSP packing algorithm to maximize low data width operations on a single DSP block, but there is a lack of co-optimization of the hardware implementation and neural network models. In summary, there is still a lack of co-optimization framework that takes the neural network model accuracy and fine-grained FPGA implementation efficiency into consideration at the same time for FPGAs.

\begin{figure*}[t]
	\centering	\includegraphics[width=0.75\linewidth]{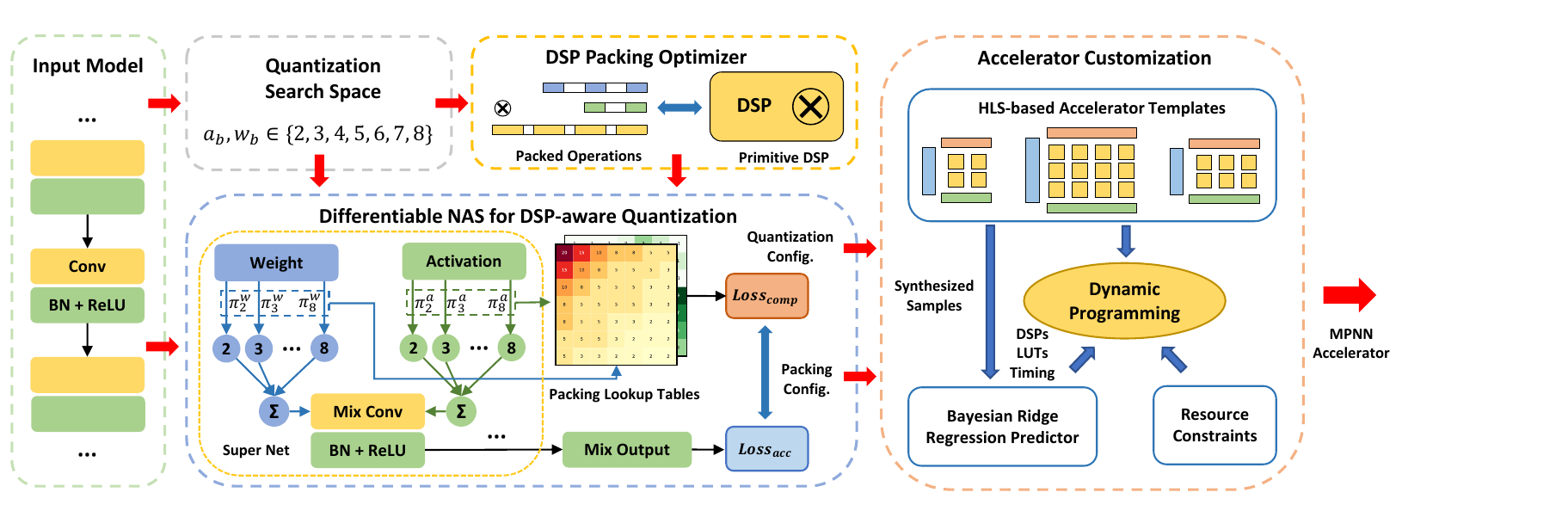}
	\caption{Overview of the proposed mixed-precision neural network accelerator design framework.}
\vspace{-1.5em}
\label{fig:overview}
\end{figure*}

\section{MPNN Accelerator Design Framework}
In this work, we propose a mixed-precision neural network accelerator design framework as shown in Fig. \ref{fig:overview}. Essentially, it is a hardware and software co-optimization framework for MPNNs and seeks to optimize both the processing performance and accuracy. In general, it adjusts the quantization of MPNNs to fulfill the accuracy requirement and optimize the resulting accelerator implementation which mainly relies on the various low data width operation mapping  efficiency over primitive DSPs within FPGAs. 

The framework starts with a floating point neural network model or fixed point model with unified quantization. With the model architecture, it utilizes differentiable NAS to determine the optimized MPNN quantization. Specifically, it defines the quantization search space in which the data width of the model in each layer ranges from 2bit to 8bit. Based on the search space, it further extends the input neural network model by adding all the possible candidate quantization branches to all the links between layers in the original input neural network and constructs a super-net for the NAS. The branches in the super-net are weighted and they can be adapted to optimize the model accuracy through back propagation. Other than the accuracy, it also has the NAS to be aware of the hardware implementation efficiency especially the DSP requirements which is usually the resource bottleneck. While the hardware implementation relies on the model quantization in each layer as well as the low data width operation mapping efficiency over primitive DSPs in FPGAs, a \emph{DSP Packing Optimizer} is utilized to pack the various low data width operations of MPNNs within primitive DSPs. Then, we have the optimal DSP packing configurations produced by \emph{DSP Packing Optimizer} stored in a lookup table such that they can be referred to immediately during NAS and utilized to co-optimize the MPNN accuracy and DSP overhead. Basically, we utilize a combined accuracy and DSP overhead loss to train the super-net where the accuracy loss is obtained through standard forward processing and DSP overhead is evaluated based on the lookup table of the DSP packing. At the end of the super-net training, the branches with highest weights will be selected as the optimized quantization configurations and DSP packing configurations accordingly. 

After the DSP-aware quantization, the framework proceeds to the accelerator customization stage. Essentially, it orchestrates the design parameters of our pipelined neural network accelerator templates based on the optimized quantization and DSP packing configurations of the neural network model for the sake of higher performance under the specified FPGA resource constraints.
Essentially, it is an resource allocation problem that allocates the hardware resources to different pipeline stages such that the implementation of each pipeline stage is optimized and the performance of the different pipeline stages are balanced at the same time. To address the constrained resource allocation problem, we have a \emph{dynamic programming} algorithm to tune the design parameters of the pipelined accelerator. The tuning procedure requires a large number of evaluation of different design options and the evaluation metrics can be hardware overhead like DSPs and timing quality, which are prohibitively expensive using standard tools from FPGA vendors. In this work, we utilize a \emph{Bayesian Ridge Regression} predictor to estimate resource utilization and the timing of each pipeline stage implementation. Although it needs additional sampling data for pre-training of the predictors, the resulting prediction models can be orders of magnitude faster and ensures rapid accelerator customization of a specific MPNN.

\section{DSP Packing Optimizer}
In order to efficiently map mixed-precision arithmetic operations onto primitive DSPs in FPGAs, we propose our \emph{DSP Packing Optimizer} which further generalizes the state-of-the-art DSP packing algorithms \cite{HiKonv}\cite{XilinxINT4}\cite{XilinxINT8} as two optional strategies and also incorporates two additional techniques for further enhancement. For a convolution operator, the optimizer traverses all possible packing configurations to find the optimal one for each bit-width combination, and stores it in lookup tables to direct quantization search and hardware customization.
\vspace{-1.5em}

\begin{figure*}[t]
	\centering	\includegraphics[width=0.75\linewidth]{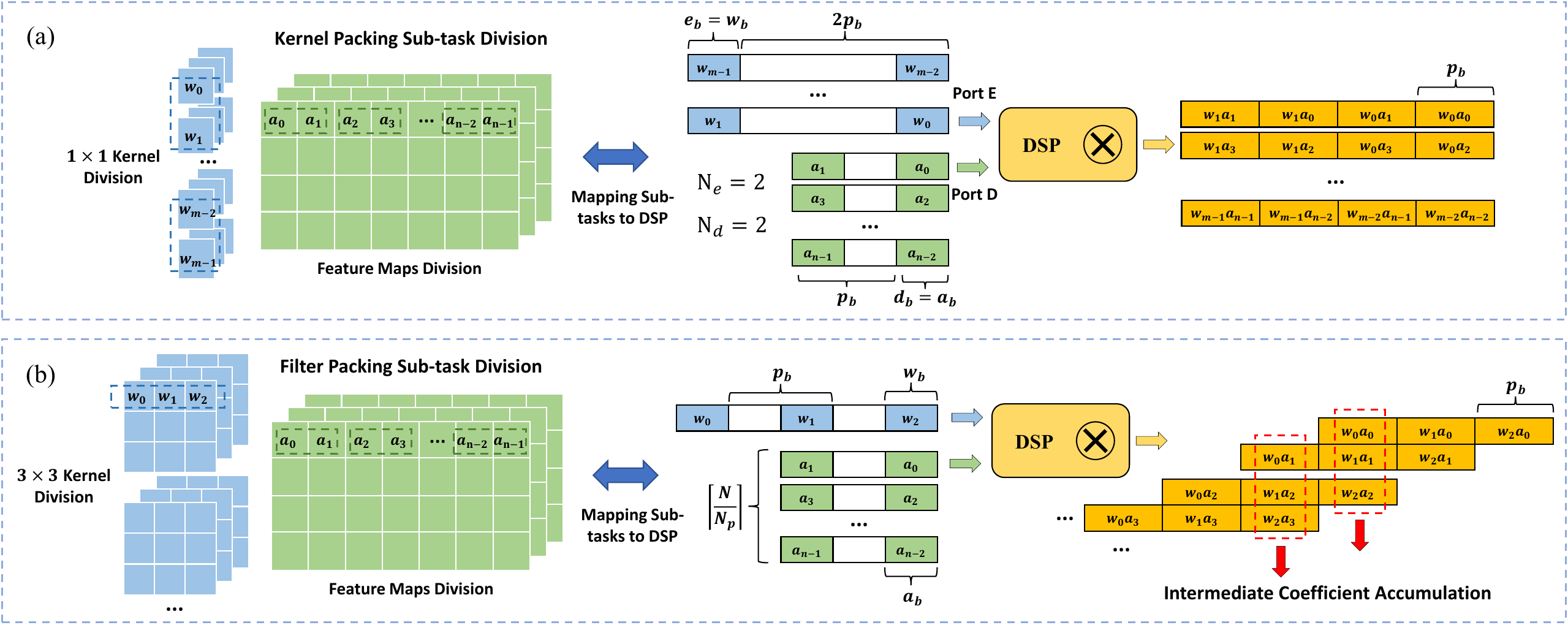}
	\caption{DSP packing strategies: (a) Kernel Packing; (b) Filter Packing. Suppose both weights and activations are unsigned.}
 \vspace{-1.5em}
	\label{fig: two packings}
\end{figure*}

\subsection{DSP Packing Strategies}
\subsubsection{Kernel Packing}
For the first strategy, as illustrated in Fig. \ref{fig: two packings}(a), weights and activations respectively from adjacent kernels and pixels are squeezed into one DSP following Eq. \ref{Eq PW-packing}. Specifically, through left-shift and addition, $N_d$ $d_b$-bit weights (activations) and $N_e$ $e_b$-bit acitvations (weights) are mapped onto two input ports, D and E, where we assume port E's bit-width is larger or equal to port D's (i.e. $P_b^E \geq P_b^D$), then $N_d \times N_e$ independent multiplications are concurrently performed within one DSP. The $g_b$ in Eq. \ref{Eq PW-packing} is called guard bits, which are deliberately preserved to prevent bit segment overlap between neighbor multiplications or to support overall result accumulation for saving decoding logic.
\vspace{-1.0em}

\begin{equation}
	\begin{split} 
		\label{Eq PW-packing}
		(\sum_{i=0}^{N_d-1}d[i]2^{i p_b})\cdot(\sum_{j=0}^{N_e-1}e[j]2^{j N_d p_b})\\
	\end{split}
\end{equation}
subject to
\begin{equation*}
	\begin{cases}
		p_b = d_b+e_b+g_b\\
		g_b \geq 0 \\
		d_b+(N_d-1)p_b \leq P_b^D\\
		e_b+(N_e-1)N_dp_b \leq P_b^E\\
	\end{cases}
\end{equation*}

\subsubsection{Filter Packing}
Instead of packing weights from adjacent kernels, \emph{Filter Packing} factorizes multidimensional convolution into 1-D counterparts, and packs the filter on an input port.
Suppose there are a $K_p$-element weight filter $f$ and a $N_p$-element activation sequence $s$. Based on the mathematical equivalence between 1-D convolution and polynomial multiplication, the convolution ($c=f * s$) can be reformulated as polynomial representation, as shown in Eq. \ref{Eq polynomial representation}, which can then be implemented with one large bit-width multiplier.
However, in practice, due to the limitation of port widths (i.e. $P_b^A$ and $P_b^W$), one multiplier cannot contain a large convolution entirely. Therefore, for processing a long $K$-element filter and a long $N$-element sequence, this strategy can be generalized by dividing the original polynomial into $\lceil \frac{K}{K_p} \rceil \times \lceil \frac{N}{N_p} \rceil$ sub-tasks. As illustrated in Fig. \ref{fig: two packings}(b), through iteratively calculating these sub-tasks and accordingly accumulating intermediate coefficients, correct convolution can be obtained. In this case, the guard bits must be greater than $\lceil \log_2min\{ K_p,N_p \} \rceil$, since $\min\{K_p,N_p \}$ accumulations are inherently introduced by its polynomial nature.
\vspace{-2.0em}

\begin{equation}
\label{Eq polynomial representation}
\begin{aligned}
        c(2^{p_b})= &f(2^{p_b})s(2^{p_b})=(\sum_{i=0}^{K_p-1} f[i] {2^{ip_b}}) \cdot (\sum_{j=0}^{N_p-1} s[j]{2^{jp_b}})\\
	=&\sum_{i=0}^{K_p+N_p-1} {c[i]}{2^{ip_b}}
\end{aligned}
\end{equation}
subject to
\begin{equation*}
\begin{cases}
	p_b=a_b+w_b+g_b\\
	g_b \geq \lceil \log_2min\{ K_p,N_p \} \rceil\\
	a_b+(N_p-1)p_b \leq P_b^A\\
	w_b+(K_p-1)p_b \leq P_b^W\\
\end{cases}
\end{equation*}

\subsubsection{Mixed Packing}
Compared with \emph{Kernel Packing} that leaves larger space (i.e. $N_dp_b$) between the operands on port E to guarantee multiplications' independence, \emph{Filter Packing} can pack operands more densely. However, for small kernel size, this advantage might not be unleashed as the filter cannot fully occupy DSP's port width (e.g. point-wise convolution). For fairly comparing different strategies and configurations, we use two metrics to evaluate our DSP packing strategies.

Primarily, we expect to maximally accommodate multiplications into these DSP primitives for highest multiplication throughput. However, we do not intuitively define it as the total number of concurrent multiplications in one DSP, because this neglects the up-rounding redundancy introduced by sub-task division in \emph{Filter Packing}, especially the division for filter whose length is shorter and can lead to severe waste. Hence, we define the multiplication throughput $T_{mul}$ as the average of the effective multiplication performed in one DSP, as shown in Eq. \ref{Eq dsp eff}. Next, based on the consideration that guard bits can be utilized for supporting accumulations before decoding, we also expect our optimizer to increase it. As \emph{Filter Packing} inherently introduces accumulations, we define extra guard bits $E_g$ with Eq. \ref{Eq DSP cascade}, and regard it as the subsidiary optimization objective.
\vspace{-1.0em}

\begin{equation}
	\label{Eq dsp eff}
	T_{mul} = 
	\begin{cases}
		N_dN_e, \ & \mbox{\emph{Kernel Packing}}\\
		\frac{KN}{\lceil\frac{K}{K_p}\rceil \lceil\frac{N}{N_p}\rceil}, \ & \mbox{\emph{Filter Packing}}
	\end{cases}
\end{equation}

\vspace{-1.5em}
\begin{equation}
	\label{Eq DSP cascade}
	E_g = 
	\begin{cases}
		g_b, \ & \mbox{\emph{Kernel Packing}}\\
		g_b - \lceil \log_2 \min\{K_p,N_p \} \rceil, \ & \mbox{\emph{Filter Packing}}
	\end{cases}
\end{equation}

\subsection{Packing Enhancement}
As explained in Eq. \ref{Eq PW-packing} and Eq. \ref{Eq polynomial representation}, multiplication throughput is restricted by both guard bits and port width constrains. In order to further enhance it, our optimizer also introduces two enhancement techniques.

\subsubsection{1-bit Overpacking}
The guard bits in packing algorithms are deliberately preserved to avoid result overlap.
Nonetheless, they inevitably consume the precious bit-widths. Inspired by \cite{DSP-Packing}, we introduce \emph{1-bit overpacking} (i.e. allowing 1-bit overlap) to mitigate this constrain, and further propose a method to fully compensate the error.

When 1-bit overlap is permitted, the LSB of high-position segment $B_{LS}^H$ and the MSB of low-position segment $B_{MS}^L$ overlaps with each other. In order to decode correctly, we recalculate $B_{LS}^H$ with additional LUT resources, and leverage it for calibration. As shown in Fig. \ref{fig: overpacking}, the LSB of a product can be obtained by applying an AND gate to the LSBs of its multiplicants. If the high-position segment is the sum of several products, we apply an XOR logic to all products' LSBs to calculate $B_{LS}^H$. For correcting $B_{MS}^L$, we directly add the recalculated $B_{LS}^H$ to it to counteract the contamination. For high-position segment, the error is possibly introduced by the sign extension of the low-position segment, which is equivalent to subtracting the result by one. To detect and compensate the unknown extension, we add the XOR of the overlapped bit and $B_{LS}^H$ for correction.

\begin{figure}[htbp]
	\centering
	\includegraphics[width=0.55\linewidth]{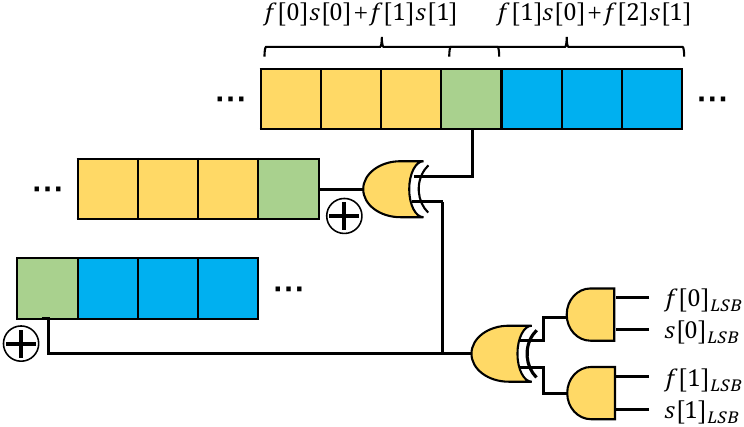}
	\caption{1-bit overpacking correction.}
	\label{fig: overpacking}
\vspace{-1.5em}
\end{figure}

\subsubsection{Operand Separation}
In packing algorithms, we expect to find suitable $K_p$ and $N_p$ (or $N_d$ and $N_e$) combinations that can fully occupy the precious port widths. However, as these values are discrete, it's unavoidable to leave some bit-width unused. This phenomenon is especially conspicuous for large bit-width combinations, since the available packing choices are more limited. Therefore, we propose using \emph{Operand Separation} to alleviate it.

The basic idea is to split a high bit-width operand (weight or activation) into two low bit-width counterparts. Take \emph{Filter Packing} as an example. Following Eq. \ref{Eq high-low sep}, a $w_b$-bit filter $f[k]$ can be split into a $(w_b - \lceil w_b/2 \rceil)$-bit $f_H[k]$ and a $(\lceil w_b/2 \rceil)$-bit $f_L[k]$. Then, the original polynomial multiplication can be reformulated as two low bit-width counterparts, and be separately implemented with two multipliers. Clearly, the separation halves the number of concurrent multiplications in one DSP. However, the resulting smaller bit-widths might be able to occupy input ports more densely, thus might serve to further enhance multiplication throughput.
\vspace{-0.5em}

\begin{equation}
	\label{Eq high-low sep}
	\begin{split}
		c(2^{p_b}) &= \sum_{i=0}^{K_p-1}\sum_{j=0}^{N_p-1} f[i]s[j]{2^{(i+j)p_b}}\\
		&=\sum_{i=0}^{K_p-1}\sum_{j=0}^{N_p-1} (f_H[k]2^{\lceil \frac{w_b}{2} \rceil} + f_L) s[j]{2^{(i+j)p_b}}\\
		&=2^{\lceil \frac{w_b}{2} \rceil}f_H(2^{p_b})s(2^{p_b})+f_L(2^{p_b})s(2^{p_b})
	\end{split}
\end{equation}

\section{DSP-aware Quantization}
Inspired by prior works \cite{EDD}\cite{Rethinking}, we mainly leverage the differentiable NAS technique for hardware-efficient quantization of MPNNs. 
As DSP is usually the resource bottleneck of NN accelerators on FPGAs \cite{DNNBuilder}\cite{DSP-Packing}\cite{EDD}, it is utilized as the major hardware metric in the hardware-aware NAS.

Given input model $\mathcal{A}$, a super-net with candidate quantization branches as shown in Fig. \ref{fig:overview} will be generated. Each branch is assigned an architecture parameter $\pi_i$ to represent selection probability. The accuracy and hardware evaluation metrics are then formulated as differentiable functions with respect to these parameters.
For inference accuracy, each layer's weight and activation candidate branches are weighted by these selection parameters in propagation such that the mixed accuracy loss ${Loss}_{acc} (\pi^w,\pi^a)$ is calculated.
For hardware-efficiency, to fully exploit the critical DSP resources on FPGAs, we regard the packed multiplications as one DSP operation, and propose to use the total DSP operations of all layers for evaluation, as defined in Eq. \ref{eq:dsp_ops}, where $Op^l_{mul}$, $Q_w^l$, and $Q_a^l$ denote the number of multiplication operations and quantization bit-widths in the $l^{th}$ layer respectively.
To orchestrate it with the super-net, each layer's multiplication throughput lookup table $T_{mul}^l$ is weighted with the corresponding selection probability as shown in Eq. \ref{eq:dsp_eff}.
The complexity loss $Loss_{comp}(\pi^w,\pi^a)$ is therewith defined as the 
probability expectation of the total DSP operations, and is added to accuracy loss with an adjustable hyper-parameter $\eta$ for tuning relative significance as shown in Eq. \ref{eq:target2}.
In back-propagation, the two loss functions are minimized based on target dataset until converge or reaching certain epochs, through which the bit-width selections are consequently optimized. In the end, the sub-path with highest selection probability is chosen as the finalized quantization configurations.
\vspace{-0.5em}

\begin{equation}
	\label{eq:dsp_ops}
	Op_{dsp} = \sum_{l=1}^{L} \frac{Op^l_{mul}}{T^l_{mul}(w_b^l, a_b^l)}
\end{equation}

\vspace{-0.5em}
\begin{equation} \label{eq:dsp_eff}
	\overline{T_{mul}^l}(\pi^w,\pi^a)=\sum_{w_b\in Q_w}{\sum_{a_b\in Q_a}{\pi_i^w \pi_j^a T_{mul}^l (w_b, a_b)}}
\end{equation}

\begin{equation} \label{eq:loss_comp}
	Loss_{comp}(\pi^w,\pi^a) = \sum_{l=1}^{L}{\frac{{Op}_{mul}^l}{\overline{T_{mul}^l}(\pi^w,\pi^a)} \ }
\end{equation}

\vspace{-0.5em}
\begin{equation} \label{eq:target2}
Loss(\pi^w,\pi^a) = {Loss}_{acc} (\pi^w,\pi^a)+\eta Loss_{comp}(\pi^w,\pi^a)
\end{equation}

\section{Accelerator Customization}
We use HLS to design hardware templates based on our DSP packing algorithms. After quantization search, the framework will automatically configure the templates and map each layer as a pipeline stage. Each stage will be assigned $Pf^l$ DSPs to construct a parallel computing array similar with \cite{DNNBuilder}. However, as LUTs can also be efficient computing resources when bit-width is small \cite{HAO}\cite{EDD}, our design also provides an alternative to construct computing arrays with equivalent LUT arithmetic, but it must satisfy timing constrain. Given the maximum DSPs $R_{dsp}^{max}$ and LUTs $R_{lut}^{max}$, we formulate the accelerator customization problem as Eq. \ref{eq:pipeline tuning}. Here, as pipeline stages are connected through FIFOs, we assume the overall Worst Negative Slack (WNS) is determined by the worst stage.


To solve above problem, we randomly sample and synthesize a set of possible hardware configurations with our templates, and pre-train a \emph{Bayesian Ridge Regression} model to estimate each stage's DSPs $\hat{R}_{dsp}^l$, LUTs $\hat{R}_{lut}^l$, and WNS $\hat{t}_{wns}^l$.
Next, we propose utilizing \emph{dynamic programming} to find the optimal resource allocation, as shown in Algorithm \ref{dynamic programming}.
It employs a three-dimension table to memorize the optimal pipeline configurations of a sub-problem, and leverages recurrence relation to solve a larger sub-problem.
Specifically, we use $Lat[l][R_{dsp}^{cur}][R_{lut}^{cur}]$ to represent the minimal latency when $R_{dsp}^{cur}$ DSPs and $R_{lut}^{cur}$ LUTs are available for the first $l$-stage pipeline of the entire design. If $\hat{R}_{dsp}^l$ out of the $R_{dsp}^{cur}$ DSPs and $\hat{R}_{lut}^l$ out of the $R_{lut}^{cur}$ LUTs are allocated for the $l^{th}$ stage, the recurrence relation can be formulated as Eq. \ref{eq:recurrence}. Based on this, we can search for the optimal solution in bottom-up fashion. The time complexity of the proposed algorithm only grows linearly with the depth and resource scale, which means a thorough design space exploration is completely affordable.
\vspace{-1.5em}

\begin{equation}
	\begin{aligned}
		\min\quad &Lat = \max \{Lat^l\} = \max \{ \frac{Op_{dsp}^l}{Pf^l}\} \\
		s.t.\quad &\sum\limits_{l=1}^{L} R_{dsp}^l \leq R_{dsp}^{max}, \ \sum\limits_{l=1}^{L} R_{lut}^l \leq R_{lut}^{max}\\
		&\min \{t_{wns}^l\} > 0\\
	\end{aligned}
	\label{eq:pipeline tuning}
\end{equation}


\begin{equation}
\label{eq:recurrence}
    \begin{aligned}
 Lat[l][R_{dsp}^{cur}][R_{lut}^{cur}]= \max \{\frac{Op_{dsp}^l}{Pf^l}, Lat^{pre}\}
    \end{aligned}
\end{equation}
where
\begin{equation*}
	\begin{aligned}
            Lat^{pre} &= Lat[l-1][\hat{R}_{dsp}^{pre}][\hat{R}_{lut}^{pre}]\\
		\hat{R}_{dsp}^{pre} &= R_{dsp}^{cur} - \hat{R}_{dsp}^l\\
		\hat{R}_{lut}^{pre} &= R_{lut}^{cur} - \hat{R}_{lut}^l 
	\end{aligned}
\end{equation*}

\begin{algorithm}[t]
	\caption{Accelerator Customization}
	\label{dynamic programming}
	Initialize all $Lat=\max \{ {Op_{dsp}^l} \}$ and Config.\\
	\For{$l=1, 2, \dots, L$}{
		\For{$R_{dsp}^{cur}=0$ to $R_{dsp}^{max}$}{
			\For{$R_{lut}^{cur}=0$ to $R_{lut}^{max}$}{
				\For{all possible $Pf^l$}{
					Estimate $\hat{R}_{dsp}^l$, $\hat{R}_{lut}^l$, and $\hat{t}_{wns}^l$\\
					$Lat^{new}=\max \{\frac{Op_{dsp}^l}{Pf^l},Lat^{Pre}\}$\\
					$C1 = Lat^{new} < Lat[l][R_{dsp}^{cur}][R_{lut}^{cur}]$\\
					$C2 = \hat{R}_{dsp}^l \leq R_{dsp}^{cur}$ \& $\hat{R}_{lut}^l \leq R_{lut}^{cur}$ \\
					$C3 = \hat{t}_{wns}^l > 0$\\
					\If{$C1 \& C2 \& C3$}{update $Lat[l][R_{dsp}^{cur}][R_{lut}^{cur}]$ and corresponding Config.}
				}
			}
		}
	}
Return $Lat[L][R_{dsp}^{max}][R_{lut}^{max}]$ and Config.
\end{algorithm} 


\section{Experiment}
\subsection{Experiment Setup}
We evaluate our framework with two datasets, the single object detection dataset in Design Automation Conference System Design Contest (DAC-SDC) \cite{DAC-SDC} and CIFAR-10. For DAC-SDC, we deploy two commonly-adopted models, UltraNet \cite{UltraNet} and SkyNet\cite{SkyNet}, with full pipeline structure, and target the bit-width selection from top-3 teams as the baseline. In order to fairly compare with previous results, we retrain and evaluate all models with the same hyper-parameter on the DAC-SDC public dataset. For CIFAR-10, we adopt a VGG-alike model (denoted as VGG-Tiny) with 6 convolution layers and one fully-connected layer, and manually design the bit-width selection as the baseline. 
For deployment, we target an embedded FPGA Ultra96-V2 (with 360 DSPs), and measure throughput, inference accuracy, utilization, and energy consumption for evaluation. In experiments, test frames are loaded into the DDR in advance and are then transferred to the accelerator by DMA. The throughput and energy consumption during inference are then measured following DAC-SDC 2022\footnote{\url{https://github.com/jgoeders/dac_sdc_2022}}. The inference accuracy is calculated with Intersection-Over-Union (IOU) for DAC-SDC and top-1 accuracy for CIFAR-10.
\vspace{-0.35em}



%


\subsection{DSP Packing Efficiency Comparison}
We start with our \emph{DSP Packing Optimizer}. As an illustrative example, Fig. \ref{fig: 3x3 comparison} compares a $3 \times 3$ kernel's multiplication throughput lookup tables that are respectively searched by HiKonv\cite{HiKonv} and our optimizer. As can be seen, our optimizer achieves higher multiplication throughput in 25 out of the 49 combinations. Furthermore, for $1 \times 1$ and $5 \times 5$ kernels, 16 and 27 cases witness throughput improvement respectively. As for the LUT overhead, we randomly sample and synthesize 30 optimized combinations. The results indicate that only 16.4 extra LUTs are required on average. These results demonstrates the effectiveness of our \emph{DSP Packing Optimizer} in generating more efficient DSP packing configurations.
\vspace{-0.7em}

\begin{figure}[htbp]
	\centering
	\includegraphics[width=0.8\linewidth]{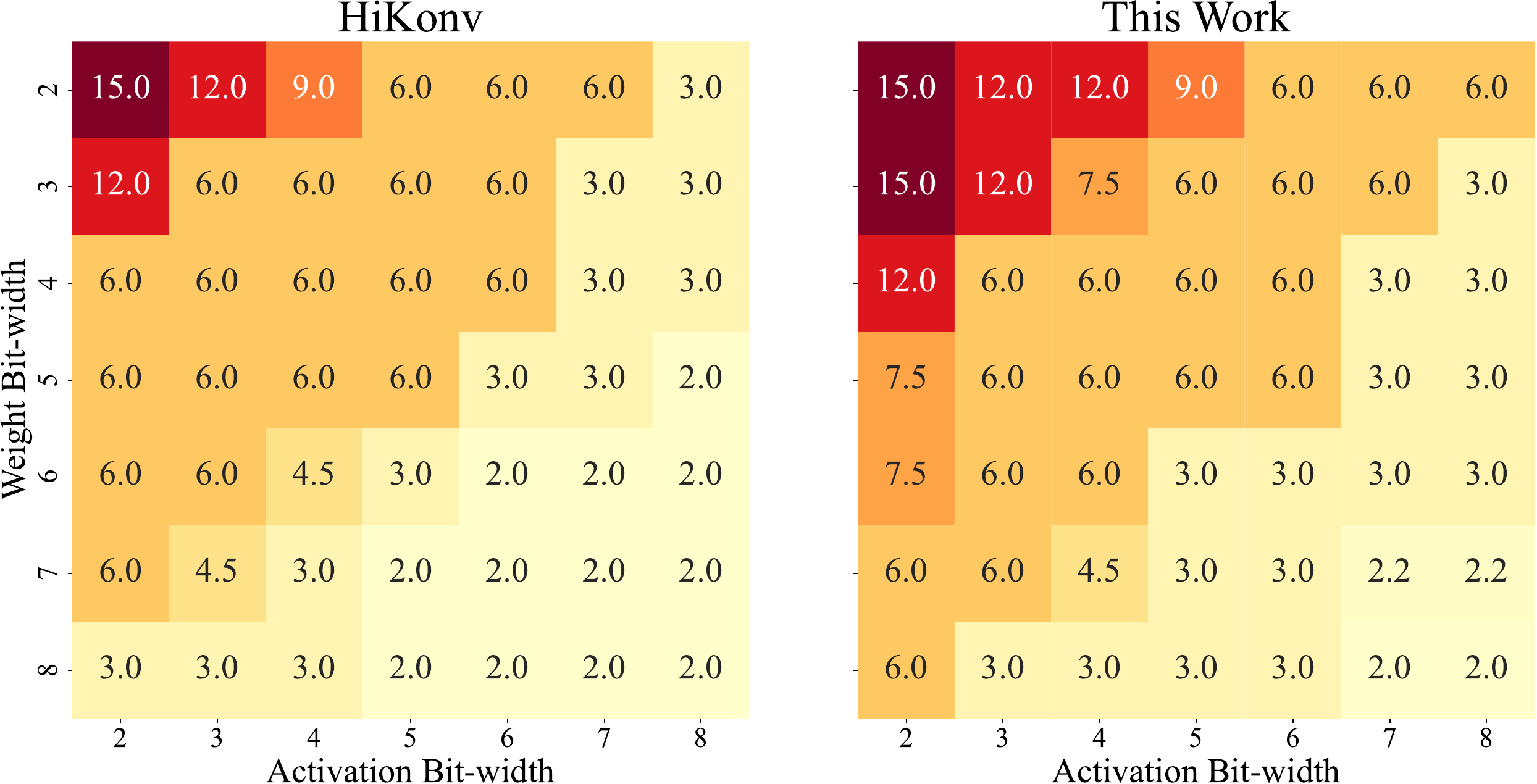}
    \setlength{\abovecaptionskip}{-0.1em}
	\caption{DSP packing efficiency comparison for a $3 \times 3$ convolution kernel.}
 \vspace{-1.0em}
	\label{fig: 3x3 comparison}
\end{figure}

\begin{figure}[htbp]
	\centering
	\includegraphics[width=0.5\linewidth]{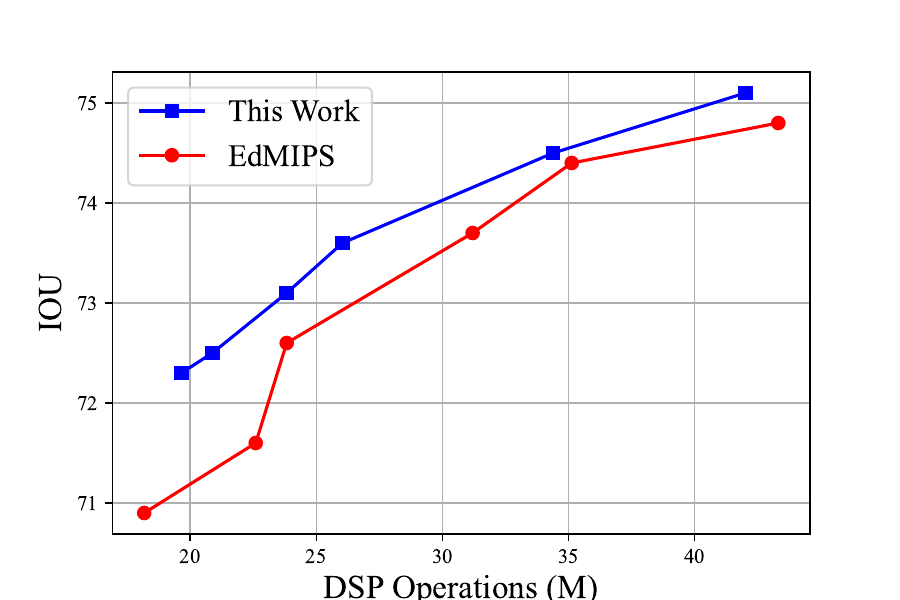}
    \setlength{\abovecaptionskip}{-0.1em}
    \caption{DSP-aware NAS results comparison with EdMIPS\cite{Rethinking}.}
	\vspace{-1.3em}
	\label{fig: NAS comparison}
\end{figure}


\subsection{Bit-width Search Results}

\begin{figure*}[thbp]
	\centering
	\includegraphics[width=0.8\linewidth]{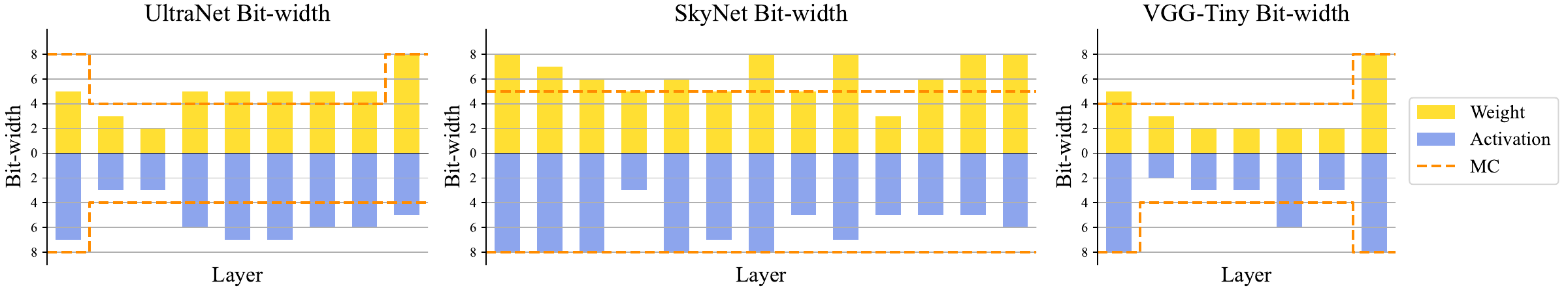}
	\caption{Bit-width selections of our mixed-precision models and manually crafted counterparts (MC).}
	\label{fig: NAS results}
	\vspace{-1.0em}
\end{figure*}

\subsubsection{NAS Comparison}
To assess our DSP-aware quantization search, we compare it with EdMIPS\cite{Rethinking} on UltraNet. Following Eq. \ref{eq:dsp_ops}, we target DSP operations as the proxy signal to balance inference accuracy and DSP operations, while EdMIPS uses the product of activation bit-width and weight bit-width to formulate computation complexity loss. Through adjusting hyper-parameter $\eta$, different solutions can be obtained as the relative significance changes, as illustrated in Fig. \ref{fig: NAS comparison}. Clearly, our method produces a pareto-optimal curve with respect to both accuracy and DSP operations, which demonstrates our metrics can effectively direct the NAS to conduct DSP-aware quantization.

\subsubsection{Bit-width Selection Comparison}
Then, we compare the searched bit-width settings with the manually crafted counterparts in Fig. \ref{fig: NAS results}. For UltraNet, we compare with iSmart\footnote{\url{https://github.com/jgoeders/dac_sdc_2021_designs/tree/main/iSmart}} (2nd in DAC-SDC 2021). They adopt high bit-width in the first and last layers, and apply 4-bit weight and activation in the rest layers such that they can leverage one DSP to pack 6 multiplications. Similar quantization scheme is also applied to VGG-tiny. For SkyNet\footnote{\url{https://github.com/jgoeders/dac_sdc_2021_designs/tree/main/SkrSkr}}, we quantize the weight and activation to 5 bits and 8 bits by referring to SkrSkr (1st in DAC-SDC 2021), and squeeze two multiplications into one DSP.

The experiment results demonstrate that our NAS can effectively optimize both inference accuracy and DSP operations. For accuracy, mixed-precision UltraNet, SkyNet, and VGG-Tiny respectively achieve IOU or top-1 accuracy of $74.29\%$, $75.44\%$ and $91.36\%$, while the accuracy of handcrafted designs are $73.37\%$, $74.85\%$ and $91.45\%$. Except for a trivial loss for VGG-Tiny, the other two models witness non-trivial accuracy improvement. As for computation complexity, $27.12\%$, $44.10\%$ and $42.71\%$ of the DSP operations are reduced.


The optimization can be explained as our NAS conducts a more reasonable bit-width allocation for each layer based on its sensitivity and computation complexity. The second and third layer of UltraNet dominate nearly $60\%$ of the overall MACs. Hence, our NAS adopts ultra-low bit-width combinations to pack 12 multiplications onto one DSP block, and in the meanwhile, increases the bit-widths in behind layers to counteract the accuracy loss. SkyNet mainly consist of six stacked bundles of depth-wise and point-wise convolution. Due to the huge complexity dichotomy between these two types of operators, larger bit-width is preferred for depth-wise convolution. In addition, unlike SkrSkr that uniformly applies larger bit-width to activation than weights, our NAS takes the opposite strategy in several layers. For VGG-Tiny, our NAS nearly choose lower bit-width for every middle layer. This means we underestimate model's tolerance to quantization when manually designing the bit-width settings.
\vspace{-0.35em}

\begin{table*}[t]
	\caption{Deployment details of the mixed-precision models and manually crafted counterparts.}
	\label{deployment comparison}
	\centering
	\begin{tabular}{ccccccccccc}
		\hline
		Backbone                                       & Implem.                           & Accuracy                                      & $Op_{dsp}$ (M)                                    & $Pf_{dsp}$ & $Pf_{lut}$                    & DSP                   & kLUT          & BRAM           & FPS              & Power (W) \\ \hline
		\multicolumn{1}{c|}{\multirow{4}{*}{UltraNet}} & \multicolumn{1}{c|}{MC-HP}     & \multicolumn{1}{c|}{73.37\%}                  & \multicolumn{1}{c|}{40.78}                   & 268   & \multicolumn{1}{c|}{0}   & 339 (94.2\%)          & 38.0 (53.9\%) & 96.0 (44.4\%)  & 1232.10          & 1.38      \\ \cline{2-11} 
		\multicolumn{1}{c|}{}                          & \multicolumn{1}{c|}{Mix-BP}     & \multicolumn{1}{c|}{\multirow{3}{*}{74.29\%}} & \multicolumn{1}{c|}{\multirow{3}{*}{29.72}}  & 196   & \multicolumn{1}{c|}{0}   & \textbf{239 (66.4\%)} & 38.2 (54.2\%) & 108.5 (50.2\%) & 1234.59          & 1.39      \\
		\multicolumn{1}{c|}{}                          & \multicolumn{1}{c|}{Mix-HP}     & \multicolumn{1}{c|}{}                         & \multicolumn{1}{c|}{}                        & 296   & \multicolumn{1}{c|}{0}   & 332 (92.2\%)          & 44.6 (63.2\%) & 130.0 (60.2\%) & 1954.30          & 1.44      \\
		\multicolumn{1}{c|}{}                          & \multicolumn{1}{c|}{Mix-LUT} & \multicolumn{1}{c|}{}                         & \multicolumn{1}{c|}{}                        & 264   & \multicolumn{1}{c|}{128} & 307 (85.3\%)          & 62.0 (87.9\%) & 160.5 (74.3\%) & \textbf{2534.20} & 1.53      \\ \hline
		\multicolumn{1}{c|}{\multirow{4}{*}{Skynet}}   & \multicolumn{1}{c|}{MC-HP}     & \multicolumn{1}{c|}{74.85\%}                  & \multicolumn{1}{c|}{200.6}                   & 168   & \multicolumn{1}{c|}{0}   & 228 (63.3\%)          & 44.2 (62.6\%) & 215.0 (99.5\%) & 193.30           & 2.99      \\ \cline{2-11} 
		\multicolumn{1}{c|}{}                          & \multicolumn{1}{c|}{Mix-BP}     & \multicolumn{1}{c|}{\multirow{3}{*}{75.44\%}} & \multicolumn{1}{c|}{\multirow{3}{*}{112.13}} & 98    & \multicolumn{1}{c|}{0}   & \textbf{179 (49.7\%)} & 35.8 (50.8\%) & 210.5 (97.5\%) & 193.21           & 2.66      \\
		\multicolumn{1}{c|}{}                          & \multicolumn{1}{c|}{Mix-HP}     & \multicolumn{1}{c|}{}                         & \multicolumn{1}{c|}{}                        & 183   & \multicolumn{1}{c|}{0}   & 273 (75.8\%)          & 40.6 (57.6\%) & 200.5 (92.8\%) & \textbf{330.93}  & 1.72      \\
		\multicolumn{1}{c|}{}                          & \multicolumn{1}{c|}{Mix-LUT} & \multicolumn{1}{c|}{}                         & \multicolumn{1}{c|}{}                        & 119   & \multicolumn{1}{c|}{64}  & 209 (58.1\%)          & 43.6 (61.8\%) & 200.5 (92.8\%) & 330.92           & 3.02      \\ \hline
		\multicolumn{1}{c|}{\multirow{4}{*}{Vgg-Tiny}} & \multicolumn{1}{c|}{MC-HP}     & \multicolumn{1}{c|}{91.45\%}                  & \multicolumn{1}{c|}{25.78}                   & 263   & \multicolumn{1}{c|}{0}   & 278 (77.2\%)          & 42.5 (60.2\%) & 213.5 (98.8\%) & 2011.55          & 1.79      \\ \cline{2-11} 
		\multicolumn{1}{c|}{}                          & \multicolumn{1}{c|}{Mix-BP}     & \multicolumn{1}{c|}{\multirow{3}{*}{91.36\%}} & \multicolumn{1}{c|}{\multirow{3}{*}{14.77}}  & 167   & \multicolumn{1}{c|}{0}   & \textbf{178 (49.4\%)} & 32.1 (45.6\%) & 128.0 (59.3\%) & 2011.29          & 1.43      \\
		\multicolumn{1}{c|}{}                          & \multicolumn{1}{c|}{Mix-HP}     & \multicolumn{1}{c|}{}                         & \multicolumn{1}{c|}{}                        & 333   & \multicolumn{1}{c|}{0}   & 348 (96.7\%)          & 46.6 (66.0\%) & 133.5 (61.8\%) & 3970.87          & 1.36      \\
		\multicolumn{1}{c|}{}                          & \multicolumn{1}{c|}{Mix-LUT} & \multicolumn{1}{c|}{}                         & \multicolumn{1}{c|}{}                        & 307   & \multicolumn{1}{c|}{144} & 340 (94.4\%)          & 62.7 (89.0\%) & 158.0 (73.0\%) & \textbf{4877.76} & 1.44      \\ \hline
	\end{tabular}
\vspace{-1.0em}
\end{table*}

\begin{table*}[]
\caption{Hardware comparison with previous co-design works.}
	\label{Comparision with previous works}
	\centering
	\begin{tabular}{cccccccc}
		\hline
		& FILM-QNN \cite{FILM-QNN}                   & N3H-Core \cite{N3H-Core}                    & HAO \cite{HAO}        & \multicolumn{1}{c|}{SEUer \cite{SEUer}}     & \multicolumn{3}{c}{This Work}         \\ \hline
		DSP Optimization                                                     & INT4, INT8                  & Bit-Parallel                 & INT8        & \multicolumn{1}{c|}{HiKonv}    & \multicolumn{3}{c}{Packing Optimizer} \\
		Backbone                                                             & ResNet-50                   & ResNet-18                    & Searched    & \multicolumn{1}{c|}{UltraNet}  & UltraNet    & SkyNet     & VGG-Tiny   \\
		Precison                                                             & 95\%W4+5\%W8, A5            & Mix(2-8)                     & W-mixed, A8 & \multicolumn{1}{c|}{4W4A, 8FL} & Mix(2-8)    & Mix(2-8)   & Mix(2-8)   \\ \hline
		\multicolumn{1}{c|}{Platform}                                        & \multicolumn{1}{c|}{ZCU102} & \multicolumn{1}{c|}{XC7Z045} & \multicolumn{5}{c}{ZU3EG}                                                            \\ \hline
		DSP                                                                  & 2092                        & 900                          & 360         & \multicolumn{1}{c|}{337}       & 307         & 273        & 340        \\
		kLUT                                                                 & 180.1                       & 152.9                        & 55.1        & \multicolumn{1}{c|}{50.2}      & 62.0        & 40.6       & 62.7       \\
		Frequency(MHz)                                                       & 150                         & 100                          & -           & \multicolumn{1}{c|}{250}       & 250         & 250        & 250        \\
		FPS                                                                  & 109.10                      & 123.2                        & 50.0        & \multicolumn{1}{c|}{2084.6}    & 2534.2      & 330.9      & 4877.8     \\ \hline
		GOPS                                                                 & 891.4                       & 446.8                        & 217.1       & \multicolumn{1}{c|}{828.6}     & 1007.4      & 263.2      & 1489.6     \\
		GOPS/DSP                                                             & 0.426                       & 0.50                         & 0.60        & \multicolumn{1}{c|}{2.46}      & 3.28        & 0.96       & 4.38       \\
		GOPS/kLUT                                                            & 4.948                       & 2.92                         & 3.94        & \multicolumn{1}{c|}{16.51}     & 16.25       & 6.48       & 23.76      \\ \hline
		Power(W)                                                             & 12.9                        & -                            & 5.5         & \multicolumn{1}{c|}{1.6}       & 1.5         & 1.7        & 1.4        \\
		\begin{tabular}[c]{@{}c@{}}Energy Efficiency\\ (GOPS/W)\end{tabular} & 69.1                        & -                            & 39.5        & \multicolumn{1}{c|}{533.2}     & 658.8       & 153.0      & 1099.6     \\ \hline
	\end{tabular}
\vspace{-1.0em}
\end{table*}

\subsection{Deployment Results}

Then, we deploy these models at different resources constrains to separately compare utilization and throughput.
We firstly allocate DSPs as many as possible to deploy the manually crafted models at their highest throughput (MC-HP) that can be supported by the platform. Next, for evaluating utilization, we deliberately reduce the available DSPs to deploy our MPNNs at the nearest throughput (Mix-BP) with the corresponding baselines. Finally, we allow our framework to maximize DSP allocation (Mix-HP). Additionally, we also enable LUT-replacement (Mix-LUT) to further improve performace. All deployment results are summarized in Table \ref{deployment comparison}.

As can be observed, due to the reduced DSP operations, our mixed-precision models demand less parallel factors to achieve the same FPS. Compared with MC-HP, Mix-BP implementations theoretically reduce 72 (26.9$\%$), 70 (41.7$\%$) and 96 (36.5$\%$) parallel factors respectively. Since DSPs can be mapped to other logic (e.g. batch normalization) as well, the final DSP savings are 100 (29.5$\%$), 49 (21.5 $\%$), and 100 (36.0 $\%$).
In terms of throughput, Mix-HP implementations achieve 1.59$\times$, 1.71$\times$ and 1.97$\times$ speedup. For UltraNet and VGG-Tiny, the highest throughput is restricted by the available DSP resources, but our SkyNet designs suffer from lack of BRAMs which prevents us from utilizing more DSPs. After enabling LUT-replacement, our framework replaces 128, 64, and 144 DSPs respectively for the three MPNNs. This further boosts the FPS of UltraNet and VGG-Tiny to 2534.20 and 4877.76. For SkyNet, while FPS remains the same, we are surprised to find that power consumption is increased from 1.72 to 3.02 W after replacing a part of DSPs.

In Table \ref{Comparision with previous works}, we also compare our designs with prior FPGA-based accelerators that focus on quantization and implementation co-optimization.
FILM-QNN \cite{FILM-QNN} restricts quantization choices as either W4A5 or W8A5, and respectively applies INT4 and INT8 for packing optimization.
In order to support flexible bit-widths, N3H-Core \cite{N3H-Core} implements 4-bit multiplication with DSPs, and applies bit-serial LUT-cores for others. Similarly, HAO \cite{HAO} leverages INT8 for high-precision multiplications and LUTs for those under 4-bit. The champion team of DAC-SDC 22, SEUer \cite{SEUer}, adopts the same bit-width and packing strategy as our UltraNet baseline, but they further boost the throughput by replacing the high bit-width arithmetic with LUTs in the first and last layer, which leads to poor WNS as a result.
Based on optimized DSP packing strategies, DSP-aware bit-width exploration, and fine-grained resource allocation scheme, our solutions show unparalleled throughput and arithmetic intensity compared with these designs.

\section{Conclusion}
In this paper, we present a framework to co-optimize the implementation and quantization of MPNNs on FPGAs. We further optimize the state-of-the-art DSP packing algorithms to support efficient implementation of arbitrary-precision convolution arithmetic. Then, we leverage differentiable NAS for automatically crafting bit-width settings based on a comprehensive assessment of accuracy and DSP operations. Finally, with our fine-grained resources allocation scheme, pipelined accelerators are customized according to specific resource requirements. According to our experiment results, our solutions reveal superior accuracy, resource utilization, and throughput, compared with manually crafted solutions and other related designs.

\newpage
\bibliographystyle{IEEEtran}
\bibliography{myreferences}

%

\end{document}